\documentclass[12pt]{iopart}

\usepackage{graphicx}

\begin{document}

\title{FSQP nuclear matrix elements for two neutrino double beta decays}

\author[first]{H. Ejiri}
\address{1. Research Center for Nuclear Physics, 
Osaka University, Ibaraki, Osaka, 567-0047, Japan 
}
\ead{ejiri@rcnp.osaka-u.ac.jp}

\vspace{10pt}
\begin{indented}
\item[] 
\end{indented}

\begin{abstract}
Nuclear matrix elements (NMEs $M^{2\nu}$) for two neutrino double beta decays (DBDs) are discussed 
in terms of the Fermi Surface Quasi Particle model (FSQP). The NMEs for  
$0^+\leftrightarrow 0^+$ ground-state-to-ground-state DBDs depend on the Fermi surface 
shell configuration  and the nuclear core polarization.  
The evaluated NMEs $M^{2\nu}(FSQP)$ reproduce well the observed NMEs $M^{2\nu}(EXP)$ for 2$\nu \beta ^-\beta ^-$ decays.  
 The NMEs $M^{2\nu}(FSQP)$ for the  2$\nu$ DBDs of $^{78}$Kr, $^{106}$Cd and $^{130}$Ba and $^{110}$Pd are evaluated on the basis of FSQP.
 The dependence of $M^{2\nu}(FSQP)$ and $M^{2\nu}(EXP)$ on the shell configuration is found to be seen in theoretical  NMEs for neutrinoless DBDs.
  Impacts of $M^{2\nu}(FSQP)$ on 0$\nu \beta \beta $ NMEs and 0$\nu \beta \beta $ experiments are discussed.\\


\end{abstract}

\section{Introduction}

 Neutrino-less double beta decays (0$\nu \beta \beta $ ), 
which are beyond the standard electro-weak model (SM), are unique probes for studying the Majorana nature of neutrinos ($\nu$), 
 the absolute $\nu $-mass scales, and others beyond SM. 
Nuclear matrix elements (NMEs) $M^{0\nu}$ for 0$\nu \beta \beta $ 
are crucial to  extract neutrino properties from double beta decay (DBD) experiments,
and even to design DBD detectors since the detector sensitivity depends much on $M^{0\nu}$.  
They are discussed in recent review articles \cite{eji00,eji05,avi08,ver12,ver16} and 
references therein. 

At present, the 0$\nu \beta \beta $ rates and the $\nu$-mass are not known experimentally, and thus $M^{0\nu}$ is not known experimentally. 
On the other hand,  two neutrino double beta decays (2$\nu \beta \beta $), which are within SM,
are measured experimentally for 2$\beta ^-$ DBD nuclei of current interests, and thus NMEs $M^{2\nu }$ for them are known experimentally.  

Extensive theoretical works have been made on $M^{0\nu}$ and $M^{2\nu }$.  Actually they 
are very small and sensitive to nucleonic and non-nucleonic nuclear correlations, nuclear models and nuclear structures 
 \cite{eji05,suh98,fae08,suh12,suh12b}. Therefore accurate theoretical calculations for $M^{0\nu}$ and $M^{2\nu }$ are
very hard.   
 
Recently, the Fermi Surface Quasi Particle model (FSQP) based on experimental single-$\beta $ NMEs $M^{\pm}$ 
for low-lying (fermi surface) quasi-particle states
 is shown to reproduce well  $M^{2\nu}$(EXP) 
for $\beta ^- \beta ^-$ decays extracted from the observed 2$\nu \beta ^- \beta ^-$ decay rates.  
 No experimental rates are known for 2$\nu$ECEC, 2$\nu \beta^+$EC and 2$\nu \beta^+\beta^+$ decays, 
except a geochemical experiment for $^{130}$Ba \cite{mes01}. In fact, it is widely believed that 
the 2$\nu$  and 0$\nu$ DBD nuclear processes are so different that  their NMEs of $M^{2\nu}$ and $M^{0\nu}$ are independent of each other.
 
The purpose of the present paper is to study the nuclear structure dependence  of the FSQP NMEs $M^{2\nu }(FSQP)$, 
and to evaluate the NMEs $M^{2\nu}$ for several 
 $2\nu \beta ^-\beta ^-$ and $2\nu$ ECEC, $2\nu\beta^+$EC, $2\nu \beta ^+\beta^+$ decays of current interest. Then we discuss possible nuclear structure effects on 0$\nu \beta \beta $ 
$M^{0\nu}$  on the basis of FSQP  to see if the 2$\nu$ NMEs and 0$\nu$ DBD NMEs have some common features.\\

\section{Fermi surface quasi particle model}

Let's first evaluate the 2$\nu$ DBD NMEs. The half-life $t_{1/2}$  is given by the phase space factor $G^{2\nu}$ 
 and the NME $M^{2\nu}$ as
\begin{equation}
t_{1/2}^{-1} = G^{2\nu}[ M^{2\nu}]^2. 
\end{equation}
Here the axial vector weak coupling constant $g_A$=1.267 in unit of the vector one of $g_F$ is conventionally included in $G^{2\nu}$.
The 2$\nu \beta ^- \beta ^- $ NME for the  A($Z,N$) $\rightarrow$ C($Z+2,N-2$) ground-state to ground state 
0$^+ \rightarrow 0^+$  transition is expressed as 
\begin{equation}
M^{2\nu}= \sum_i \frac{M_i^- M_i^+}{\Delta _i},
\end{equation}
where  and $M_i^-$ and $M_i^+$ are GT NMEs for
the $\beta ^- $A(Z,N) $\rightarrow$B(Z+1,N-1) and $\beta^+ $ C(Z+2,N-2) $\rightarrow $B(Z+1,N-1) transitions 
via the $i $th state in the intermediate nucleus B(Z+1,N-1), 
and $\Delta _i$ is the energy denominator \cite{eji05,ver12}.

 Note that the 2$\nu$ NMEs for 2$\nu$ECEC, 2$\nu\beta^+$EC and 2$\nu\beta^+\beta^+$decays for A($Z,N$) $\leftarrow$ C($Z+2,N-2$) are 
expressed by the same eq.(2), but their 
 halflives are deduced from eq.(1) by using their phase space factors.

 We discuss mainly DBD NMEs in medium heavy nuclei of current interest.
GT strength distributions for the DBD nuclei have been well studied by charge exchange reactions (CER), particularly by 
the high energy-resolution  ($^3$He,t) reactions, as discussed in the review articles \cite{eji00,eji05,ver12}. They show low-lying GT states 
 with weak strength of $B(GT) \approx $0.05-0.3
and a strong $\beta ^- $ GT giant resonance (GR) with $B(GT) \approx $ 1.5 $(N-Z)$ at the high excitation region of $E$=10-15 MeV. 
There is no strong $\beta^ + $ GT GR since  $\beta ^+ $ p$\rightarrow $n 
GT transitions are blocked by the neutrons in the same shell with the protons in medium heavy nuclei. 

In the FSQP model \cite{eji05,ver12,eji09,eji12},  the 2$\nu \beta \beta $ NME is expressed as the sum of the NMEs via the intermediate FSQP states.
The quasi-particle configurations involved in the transition of A(0$^+) \rightarrow $B(1$^+)\rightarrow$C(0$^+)$
are $(J_iJ_i)_0 \rightarrow (J_ij_k)_1\rightarrow (j_kj_k)_0$, where $J_i$ and $j_k$ are the $i$ neutron and $k$ proton spins. The Fermi surface is diffused 
due to the pairing interaction. The quasi-particle  pairs to be considered are the quasi-neutron  
pairs of $(J_iJ_i)_0$ in the defused Fermi surface of A, 
and the quasi-proton pairs of $(j_kj_k)$ in the defused Fermi surface of C. 
Since the quasi-neutron $J_i$ and the quasi-proton $j_k$  are in the 0$^+$ ground states of A and C, 
the intermediate states of ($J_i,j_k$) are necessarily FSQP low-lying states in B within the energy  width around the pairing energy of a few MeV.

The FSQP GT NMEs $M_i^{\pm}$ for $\beta ^{\pm}$ is simply expressed as \cite{eji05,eji09,eji12},
\begin{equation}
M_i^{\pm} = k^{\pm}M_i^{\pm}(QP), ~~~~M_i^{\pm}(QP)= P_i^{\pm}M(J_ij_i),
\end{equation}
where $M_i^{\pm}(QP)$ is the quasi-particle NME, $k^{\pm}$ is the effective axial coupling constant in units of the axial coupling $g_A$ 
 \cite{eji00,eji78} and  $P_i^{\pm}$ is the
  the pairing correlation
coefficient for $\beta ^{\pm}$ transition, and $ M(J_ij_i)$ is the single particle $J_i \rightarrow j_i$ GT NME. Here $P_i^{\pm}$ 
stands for the reduction coefficient due to the pairing correlations and $k^{\pm}$ does for the coefficient due to 
the spin-isospin correlations and nuclear medium effects as discussed before \cite{eji00,ver12,eji78}, and also  rcently on the GT and SD $\beta $ NMEs 
\cite{eji15, eji14}.

The pairing reduction coefficients are given as 
\begin{equation}
P^-(Jj)=V_J(N)U_j(Z), ~~~P^+(Jj)=U_J(N-2)V_j(Z+2),
\end{equation}
where $V_J(N)$ and $U_j(Z)$ are the occupation and vacancy coefficients for $J$ quasi-neutron and $j$ quasi-proton orbits
 in the initial nucleus A,
and $V_j(Z+2)$ and $U_J(N-2)$ are those for $j$ quasi-proton and $J$ quasi-neutron in the final nucleus C. 
 Since the same SP NME of $M(J_ij_i)$ is involved in both the  $M_i^{-}$ and $M_i^{+}$, 
the product is positive, and thus the sum is constructive.
 The GT NMEs for the FSQP states in the low excitation region are evaluated from 
the experimental CER and/or the $\beta ^{\pm}$ rates.

The unique features of the FSQP $M^{2\nu}$ are as follow. 

\begin{itemize}
\item  The 2$\nu\beta \beta $ decays are
expressed as the successive $\beta ^{\pm}$ transitions via the low-lying FSQP intermediate states, where the single $\beta ^{\pm}$ NMEs , including $g^{eff}_A/g_A$,
are given experimentally by CERs and $\beta $/EC rates. Thus one does not need to evaluate pure theoretically the nuclear correlations and the effective  $g_A$
involved in NMEs, which are very hard.  There are no appreciable contributions from the GT giant resonance to $2\nu \beta \beta $
 NMEs as evaluated theoretically \cite{eji95}. 

\item The FSQP $M^{\pm}$ is smaller than the SP NME by the 
pairing coefficient $P^{\pm}\approx $0.45-0.25 and the effective coupling coefficient  $k^{\pm} \approx $ 0.3-0.2. \cite{eji05, ver12,eji78}. Thus $M^{2\nu}$
becomes smaller by the coefficient $k^-P^-k^+P^+ \approx $0.01-0.005 with respect to the single particle value. 

\item  The values $M^{2\nu}$ reflect the GT $\beta ^{\pm}$ strength distributions in the intermediate nucleus. They are large in nuclei where a
large GT$^{\pm}$ strength is located at the ground state with a small $\Delta_1$, as seen in the NMEs for $^{100}$Mo $^{106}$Cd and $^{110}$Pd nuclei
 
\item  $M^{2\nu}$ depends on the shell structure as the pairing coefficient $P^{\pm}$ \cite{eji09}. 
 The product $P^-_iP^+_i$ of the pairing factors is very stable in the middle of the shell, but gets small near the shell/sub-shell  closure 
because the the vacancy amplitude $U$ or the occupation amplitude $V$ gets small just before or after the shell closure. 
\end{itemize}

\section{Two neutrino DBD NMEs}

 The 2$\beta ^-\beta ^-$ nuclei to be discussed in the present paper  
are $^{76}$Ge, $^{82}$Se, $^{96}$Zr $^{100}$Mo,
  $^{110}$Pd, $^{116}$Cd, $^{128}$Te, $^{130}$Te, and $^{136}$Xe. These are interesting for 0$\nu \beta \beta $ 
studies because of the large phase volume of $G^{0\nu}$.
The 2$\nu \beta \beta $ halflives for them, except $^{110}$Pd, are known experimentally. The observed and evaluated 2$\nu $ NMEs are shown in Table 1.  

The FSQP NME for $^{76}$Ge, $^{82}$Se and $^{110}$Pd are the present values and the NMEs for $^{136}$Xe and others are from 
the previous works of ref.\cite{eji12} and ref.\cite{eji09}. The data for single beta decays and charge exchange reactions \cite{fir99,ENSDF,fye08,fre96,bla00,bla12,bla12a,gur12}
are used to evaluate the FSQP NMEs. The present values for $^{76}$Ge and $^{82}$Se  are nearly the same as the previous values \cite{eji09}. 
The observed and FSQP NMEs are in good agreement as given in Table 1. 
 The new NME of 0.145 for the $^{110}$Pd is quite large. It gives the halflife of $t_{1/2}$= 1.3 10 $^{20}$ y. Theoretical pnQRPA values 
 of 1.1-0.91 10$^{20}$ y \cite{suh11} and 1.2-1.8 10$^{20}$ y \cite{civ98} and the SSD value of 1.2 10$^{20}$ y \cite{dom05} are close to the present value. The experimental study 
 with the 10$^{20}$ y sensitivity is encouraged.

The 2$\nu$ECEC, 2$\nu \beta ^+$EC and 2$\nu \beta ^+\beta ^+$ DBDs are not well studied because of the small phase volume. Here we 
discuss three DBDs of $^{78}$Kr, $^{106}$Cd, and $^{130}$Ba. The experimental and FSQP NMEs are shown in Table 2. The $M^{2\nu}(EXP)$ limits
for $^{78}$Kr\cite{gav13} and $^{106}$Cd\cite{ruk15} are from the ECEC data and the NME for $^{130}$Ba is from geochemical method \cite{bar10}.

\begin{table}[!b]
\caption{The 2$\nu\beta ^-\beta ^-$ NMEs  for the 0$^+\rightarrow 0^+$ ground state 0$^+$ and for the first excited 0$^+$ (*) transitions. 
$Q_{\beta \beta }$ : $Q$ value. $M^{2\nu}(EXP)$: experimental NME with $a$ ref.\cite{ago15}, $b$  ref.\cite{bar10} and others ref. \cite{eji09}.
$M^{2\nu}(FSQP)$: FSQP NME with $c$ ref.\cite{eji12},$d$ the present value and others ref.\cite{eji09}.
\label{tab:ej1}}
\begin{center}
\begin{tabular}{lccc} \hline
Transition & $Q_{\beta \beta }$ MeV  & $M^{2\nu}(EXP)$ & $M^{2\nu}(FSQP)$ \\ 
$^{76}$Ge $\rightarrow ^{76}$Se  & 2.039 &   0.063$^a$  &  0.052$^d$\\
$^{82}$Se $\rightarrow ^{82}$Se  & 2.992 &   0.050     &  0.064$^d$\\
$^{96}$Zr $\rightarrow ^{96}$Mo  & 3.346 &   0.049     &  0.045 \\
$^{100}$Mo $\rightarrow ^{100}$Mo  & 3.034 &  0.126   &  0.096 \\
$^{100}$Mo $\rightarrow ^{100}$Mo*  & 1.904 &  0.102   &  0.090 \\
$^{110}$Pd $\rightarrow ^{110}$Cd  & 2.000 &  -   &  0.145$^d$ \\
$^{116}$Cd $\rightarrow ^{116}$Sn  & 2.804 &  0.070   &  0.055 \\
$^{128}$Te $\rightarrow ^{128}$Xe & 0.867 &  0.025   &  0.019 \\
$^{130}$Te $\rightarrow ^{130}$Xe  & 2.529 &  0.018   &  0.017 \\
$^{136}$Xe $\rightarrow ^{136}$Ba  & 2.467 &  0.010$^b$   &  0.012$^c$ \\

\hline
\end{tabular}
\end{center}

\medskip 
\label{tab:eji1}
\end{table} 

\begin{table}[!t]
\caption{The 2$\nu$ECEC NME  for the 0$^+\rightarrow 0^+$ ground state transition. See the caption of the table 1.
a: ref.\cite{gav13}.  $b$: ref.\cite{ruk15}. $c$: ref.\cite{bar10}. $d$: present value.
\label{tab:ej2}}
\begin{center}
\begin{tabular}{lccc} \hline
Transition & $Q_{\beta \beta }$ MeV  & $M^{2\nu}(EXP)$ & $M^{2\nu}(FSQP)$ \\ 
$^{78}$Kr $\rightarrow ^{78}$Se & 2.866 &   $\leq$ 0.34$^a$   &  0.065$^d$ \\
$^{106}$Cd $\rightarrow ^{106}$Pd  & 2.770 &  $\leq$ 0.39$^b$   &  0.11$^d$ \\
$^{130}$Ba $\rightarrow ^{130}$Xe  & 2.610 &  0.105 $^c$   &  0.067$^d$ \\
\hline
\end{tabular}
\end{center}

\medskip 
\label{tab:eji2}
\end{table}

DBDs listed in Tables 1 and 2 are classified into three groups as given in Table 3.  
Group I: FSQP neutrons and protons are in the same $N$=3 major shell.  Group II: ESQP protons and 
neutrons are in the $N$=3 and $N$=4 major shells, respectively.  Group III: FSQP neutrons and protons 
are in the same $N$=4 major shell.  Group 1:$^{76}$Ge, $^{78}$Kr, $^{82}$Se. 
Group II: $^{96}$Zr, $^{100}$Mo, $^{106}$Cd, $^{110}$Pd, $^{116}$Cd.
Group III:  $^{128}$Te, $^{130}$Te, $^{130}$Ba, $^{136}$Xe. 

\begin{table}[htb] 
\caption{ FSQP shell configurations for the three groups (G) of the DBD nuclei. \label{tab:cd106}}
\vspace{0.5 cm}
\centering
\begin{tabular}{ccc}
\hline
G & $Z,N$ &  FSQP GT configurations  \\
I &  32$\le Z\le 36,42\le N\le$48 &   1g$^p_{9/2}$1g$^n_{9/2}$, 1f$^p_{5/2}$1f$^n_{5/2}$, 2p$^p_{3/2}$2p$^n_{1/2}$ \\
II &  40$\le Z \le$48, 56$\le N \le $68  & 1g$^p_{9/2}$1g$^n_{7/2}$ \\
III &  52$\le Z\le $56, 76$\le  N \le $82 & 2d$^p_{5/2}$2d$^n_{3/2}$, 1h$^p_{11/2}$1h$^n_{11/2}$, 3s$^p_{1/2}$3s$^n_{1/2}$\\ 
\hline
\end{tabular}
\end{table}

All FSQP states in the low excitation region are involved more or less in the 2$\nu $ DBD. In case of the group II nuclei, the FSQP protons and neutrons  are in different major shells, and 
 the ground state is the only one strong GT state of 1g$^p_{9/2}$1g$^n_{7/2}$ in the low-excitation region, as observed in the ($^3$He,t) CERs 
\cite{aki97,thi12,pup11,pup12}. 
Thus the hypothesis of the 
single state dominance (SSDM) \cite{aba84} is valid only for nuclei in this group.  In cases of the Group I and III nuclei, the  FSQP protons and neutrons   
are in the same major shell. Thus 
there  are several FSQP states in the low excitation region of the intermediate nucleus \cite{aki97,thi12,pup11,pup12}, and they all contribute to $M^{2\nu}$.

 The experimental and FSQP 2$\nu$ NMEs are shown as a function of the 
proton and neutron numbers of the intermediate nuclei in Fig.1. They show clearly similar sub-shell structures. 
The NMEs for group II nuclei get small at the opening of $Z$=40 and the closure of $Z$=50 for the g$_{9/2}$ proton shell, and get large in the middle (Fig.1.B).
The NMEs for the group III get small as the proton number approaches to the shell closure of $Z$=50 (Fig.1 C)  and as the neutron number to that of  $N$=82 (Fig.1.D).

\begin{figure}[htb]
\caption{The FSQP and experimental NMEs of  $M^{2\nu}(FSQP)$ (squares) and  $M^{2\nu}(EXP)$ (diamonds) for nuclei in the group I (A) and II (B) and 
III (C and D). 
$Z_B$ and $N_B$ are the proton and neutron numbers in the intermediate nucleus.                                                                                                                                                                                                                                                                                                                                                                                                                                                                                                                                                                                                                                                                                                                                                                                                                                                                                                                                                                                            
\label{fig:avmean}}
\begin{center}
\includegraphics[width=0.65\textwidth]{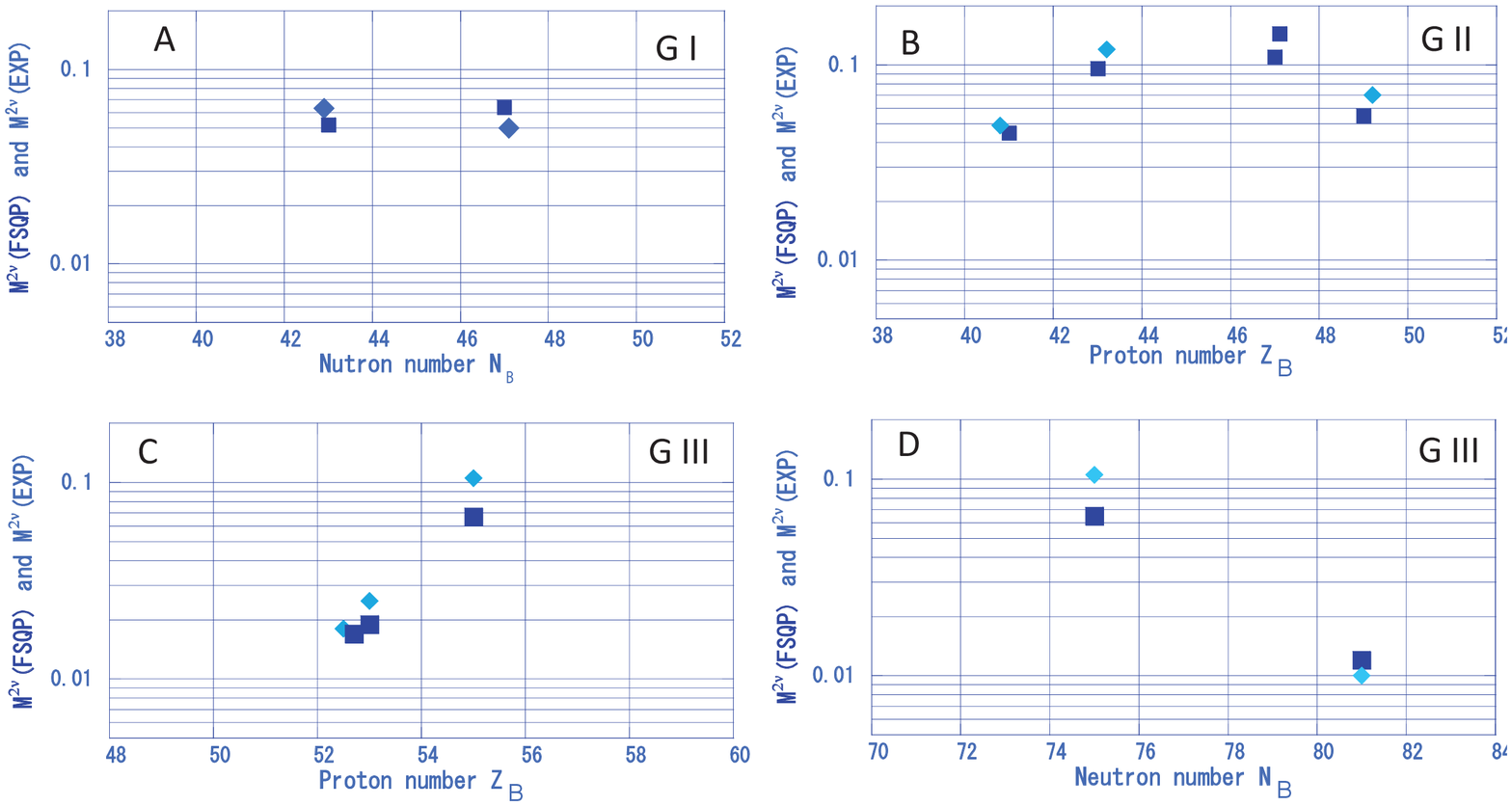}
\end{center}
\end{figure}

Thes simple systematic features of the FSQP model are different from the QRPA calculations \cite{vog86,rod06}.  The small NMEs $M^{2\nu}$ are attributed to 
 the cancellation of the  ammplitudes in the QRPA $\beta $ NMEs at a certain value of $g_{pp}$ (particle particle interaction).
In other words, the QRPA calculations reproduce  the experimental $
 M^{2\nu}$ values by adjusting $g_{pp}$ and $g_A$.  On the other hand, FSQP is based on the experimental single 
$\beta ^{\pm}$ NMEs for the low lying states, but not on any adjustable parameters.  
 
ECEC, EC$\beta ^+$ and $\beta ^+\beta ^+$  DBDs are not well studied experimentally because their phase space factors are rather small. 
The evaluated halflives on the basis of the FSQP NMEs are shown in Table 4.  
The experimental limit of 5.5 10$^{21}$y \cite{gav13} for the  $^{78}$Kr KK life  is smaller than the FSQP value by a factor around 20, 
while the limits of 2.7 10$^{20}$ y \cite{dan15} and 4.2 10$^{20}$ \cite{ruk15} for the 
$^{106}$Cd ECEC  are a factor around 15 smaller than the FSQP value. The limit of 1.1 10$^{21}$ \cite{bel16} for 2$\nu\beta ^+$EC is also 
smaller than the FSQP by a factor 20. On the other hand, 
the halflife of 2.2 10$^{21}$ y for $^{130}$Ba \cite{mes01} derived by a geochemical method is close to the FSQP life, but shorter  a little than it. 

Theoretical halflives do depend on the models and the parameters used for the calculations. 
2$\nu $ECEC halflives of 3.7 10$^{21}$ - 9.4 10$^{22}$ y for $^{78}$Kr \cite{hir94,aun96,toi97}
are shorter than the FSQP values. 
The values for $^{106}$Cd ECEC halflives are 0.12-5.5 $\times 10^{21}$ y by QRPA/RQRPA/SQRPA \cite{suh01,toi97,sto03}, 
9.7-25 $\times 10^{21}$ y by PHFB \cite{shu05} and 1.7-4.2 $\times 10^{21}$ y by SU(4) \cite{ram98}. 
The 2$\nu ECEC$ halflife for $^{130}$Ba by QRPA \cite{hir94} is 4.2 10$^{21}$ y.
On the other hand, the SSDH gives $\leq $1.6 10$^{24}$ y, $\geq $4.4 10$^{21}$ y and 5.0 10$^{22}$ y for the 2$\nu$ECEC halflives of $^{78}$Kr, 
$^{106}$Cd, and $^{130}$Ba \cite{dom05}.  The SSDH values for $^{78}$Kr and $^{130}$Ba in the group I and III are much longer than the FSQP ones because 
the calculation does not include the excited FSQP states.

\begin{table}[htb] 
\caption{ FSQP NMEs and halflives for the 2$\nu $ DBDs of $^{78}$Kr, $^{106}$Cd and $^{130}$Ba. 
$G^{2\nu}$: phase space factor from ref. \cite{suh98}.  
\label{tab:cd106life}}
\vspace{0.5 cm}
\centering
\begin{tabular}{cccc}
\hline
Decay mode & $^{78}$Kr  &    $^{106}$Cd  & $^{130}$Ba \\ 
$M^{2\nu}$ & 0.065 & 0.11 & 0.067\\
ECEC  & 1.2 10$^{23}$  &  5.2 10$^{21}$ & 5.4 10$^{21}$\\
 EC$\beta ^+ $ & 2.0 10$^{23} $ & 4.1 10$^{22}$& 1.6 10$^{23}$\\
 $\beta ^+ \beta ^+$ &  7.0 10$^{26}$  & 1.7 10$^{27}$ & 1.9 10$^{29}$\\ 
\hline
\end{tabular}
\end{table}

\section{Remarks and discussions}

Now, we discuss impact of the present discussions on the the 0$\nu \beta \beta $ NMEs $M^{0\nu }$.  
It has been believed long that i $M^{2\nu}$ is very small because it is sensitive to $g_{pp}$ and the amplitudes 
involved in $M^{2\nu}$
cancels at the appropriate value of $g_{pp}$, while  $M^{0\nu}$ is  large because it is not sensitive to $g_{pp}$ and
 ii  $M^{0\nu}$ includes several multipole NMEs and  it is not sensitive to the individual nuclear structures, 
and thus they are considered to be nearly the same for all nuclei.

On the other hand, the FSQP model shows that 
$M^{2\nu}$ is much smaller than the quasi-particle NME $M_{QP}$ by the reduction coefficient $(k^{\pm})^2 \approx $0.05-0.1 
because the observed single $\beta ^{\pm}$ GT(1$^+$) $M^{\pm}$ is smaller than the single quasi-particle $M_{QP}^{\pm}$ (GT) by the coefficient
$k^{\pm}$= $k^{eff} \approx $0.2-0.3.  The  single $\beta ^{\pm}$ SD(2$^-$) $M^{\pm}$, which is one of the major components of $M^{0\nu}$, is smaller than the single 
quasi-particle $M_{QP}^{\pm}$ (SD) by the coefficient $k^{eff}\approx $0.2-0.3 \cite{eji14}, as in case of the GT NME \cite{eji15}. Accordingly, the axial-vector component of 
$M^{0\nu}$ may be much smaller than the QP NME $M^{0\nu}_{QP}$ by the coefficient $(k^{eff})^2 \approx $ 0.05-0.1. 

The reduction coefficient is given as 
$k^{eff}=k_{\tau \sigma} \times k_{NM}$, where $k_{\tau \sigma}\approx 0.4-0.6$ and $k_{NM} \approx 0.4-0.6$ 
stand for the nucleonic $\tau \sigma $ correlation and and the non-nucleonic nuclear medium
effects \cite{eji14,eji15}. The former  is incorporated in pnQRPA with the nucleonic $\tau \sigma $ correlations, 
while the latter may be expressed in terms of the effective $g_A^{eff}/g_A^{free}$.
Actually, the values of $g_A^{eff}/g_A^{free}\approx 0.5-0.7$ are used in recent theoretical calculations such as 
shell model \cite{mar96,cau12}, pnQRPA \cite{suh13,suh14}, and IBA2 \cite{bar13}. 
Then $M^{0\nu}$ may be reduced 
with respect to the pnQRPA NME by a factor 2-3, depending on the higher multi-pole axial vector NMEs and the vector NMEs.

In the present cases of the ground-state to ground state DBDs, the values for $M^{0\nu}$ near  closed and sub-closed shells may 
get small due to the pairing factor $P^{\pm}$ as in cases of $M^{2\nu}$.  Then  both $M^{0\nu}$ and $M^{2\nu}$ for $^{136}$Xe are 
small due to the small paring factor $P^+$ near the neutron closed shell at 82. Thus the 0$\nu \beta \beta $ signal rate may get small due to the small $M^{0\nu}$, but the 
BG contribution from the high energy tail of the 2$\nu \beta \beta $ spectrum to the 0$\nu \beta \beta$ peak region gets small also 
in case of modest energy-resolution experiments.
 
\begin{figure}[htb]
\caption{Top: Average values for the calculated NMEs $M^{0\nu}$ \cite{ver16}. Bottom: The FSQP NMEs  $M^{2\nu}(FSQP)$ (squares) and  experimental NMEs 
$M^{2\nu}(EXP)$ (diamonds) for $\beta ^-\beta ^- $ DBDs.                                                                                                                                                                                                                                                                                                                                                                                                                                                                                                                                                                                                                                                                                                                                                                                                                                                                                                                                               
\label{fig:2v0n nmes}}
\begin{center}
\includegraphics[width=0.55\textwidth]{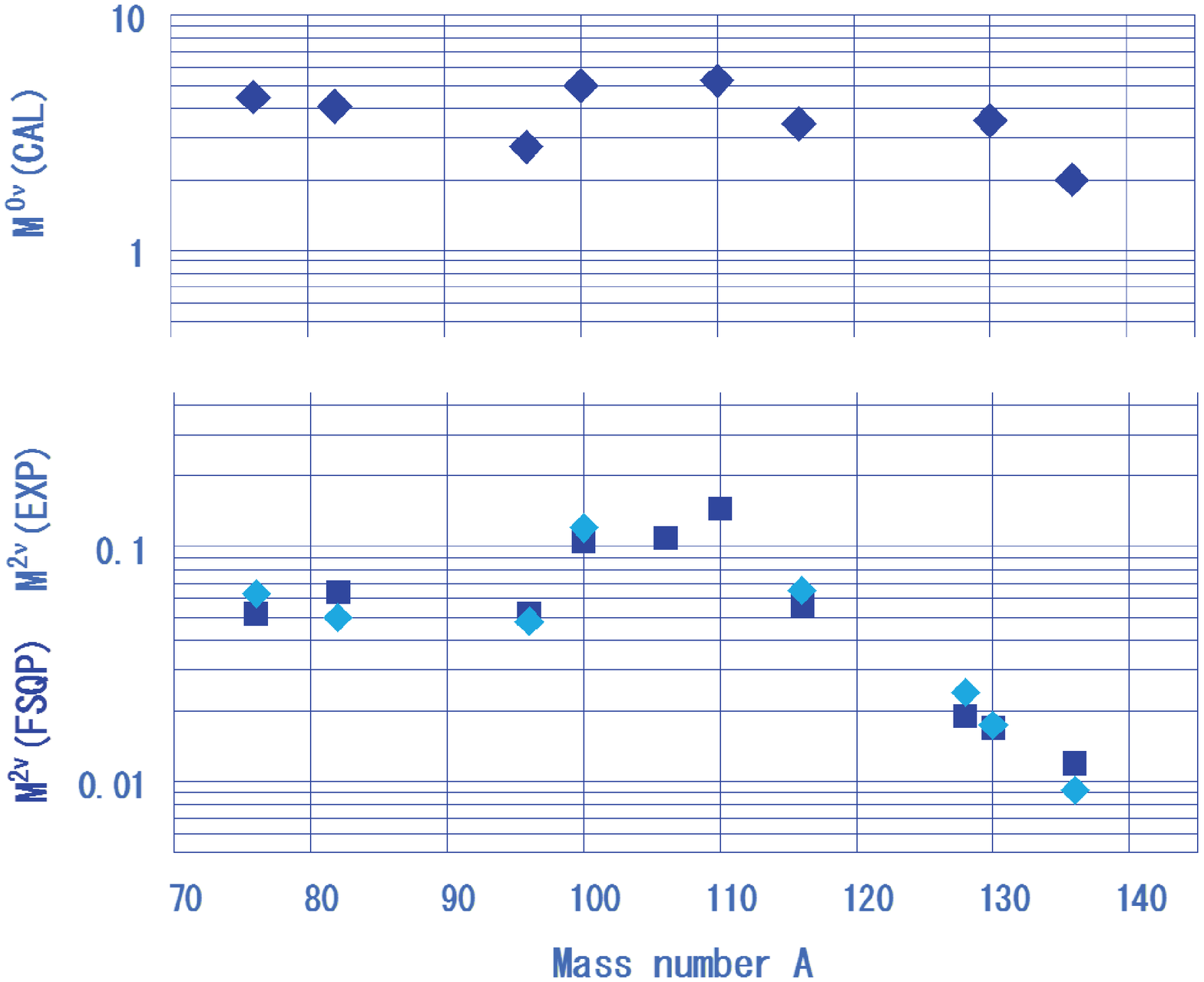}
\end{center}
\end{figure}

The 0$\nu \beta \beta $ NMEs have been calculated on various nuclei. The averaged value \cite{ver16} for each DBD isotope of the current interest is plotted in fig.2.
The 0$\nu \beta \beta $ NMEs do show the similar dependence on the shell configuration as 2$\nu \beta \beta $ NME. The 0$\nu \beta \beta $ 
NMEs get small near the sub closed shell due to the small $U$ or $V$ coefficients as in case of the 2$\nu\beta \beta $ NMES. This is in contrast to the general 
 belief that all the 0$\nu \beta \beta $ NMEs are large and same. Then, one may need to consider small and structure-dependent NMEs $M^{0\nu}$ as well in 
selecting DBD nuclei to be studied in future DBD experiments.

Experimental studies of 2$\nu$ECEC of  $^{78}$Kr, $^{106}$Cd, $^{110}$Pd, and $^{130}$Ba are encouraged to verify the FSQP prediction. 
The single $\beta ^{\pm}$ NMEs for 3$^+$ and 4$^-$ transitions in DBD nuclei are interesting to check if similar reductions 
of the NMEs as GT and SD are seen for higher multipole NMEs as well.\\

The authors thank Prof. D. Frekers and Prof. J. Suhonen for valuable 
discussions.

\end{document}